# Preparation, Characterization and electronic structure of Ti − doped $Bi_2Se_3$

Sawani Datta, Arindam Pramanik, Ram Prakash Pandeya, Anup Pradhan Sakhya, A. Thamizhavel and Kalobaran Maiti[*]

Department of Condensed Matter Physics and Material Science, Tata Institute of Fundamental Research, Homi Bhabha Road, Colaba, Mumbai 400005, India

[*]Corresponding author: kbmaiti@tifr.res.in.

**Abstract.** We report the preparation of high-quality single crystal of $Bi_2Se_3$, a well-known topological insulator and its Ti-doped compositions using Bridgeman technique. Prepared single crystals were characterized by *x*-ray diffraction (XRD) to check the crystalline structure and energy dispersive analysis of *x*-rays for composition analysis. The XRD data of Ti-doped compounds show a small shift with respect to normal $Bi_2Se_3$ indicating changes in the lattice parameters while the structure type remained unchanged – this also establishes that Ti goes to the intended substitution sites. All the above analysis establishes successful preparation of these crystals with high quality using Bridgman technique. We carried out *x*-ray photo-emission spectroscopy to study the composition via investigating the core level spectra. $Bi_2Se_3$ spectra exhibit sharp and distinct features for the core levels and absence of impurity features. The core level spectra of the Ti-doped sample exhibit distinct signal due to Ti core levels. The analysis of the spectral features reveal signature of plasmon excitation and final state satellites – a signature of finite electron correlation effect in the electronic structure.

## INTRODUCTION

The three-dimensional topological insulators (TI) are an interesting class of materials, which possesses a metallic surface states (SS) over the insulating bulk materials. The gapless linearly dispersed metallic surface states are time-reversal symmetry (TRS) protected [1]. If we break the time-reversal symmetry by suitable perturbation, a gap will open at the Dirac point of SS. Keeping this in mind, we doped Ti in a well-studied three-dimensional topological insulator, $Bi_2Se_3$ to find, whether Ti can break the TRS or not. People are investigating the electronic structure of similar systems with different kinds of magnetic elements (e.g. Fe, Cr, Mn) doping in $Bi_2Se_3$ [2,3] or any other three-dimensional topological insulator. All these dopants studied so far are the strongly correlated systems containing large 3*d* orbital occupancy and magnetic moment that enhances the level of difficulty. We chose Ti as dopant as this is expected to have one 3*d* electron for 3+ valency - usually, Ti is very stable in 4+ state, where there is no 3*d* electron. Therefore, the 3*d* level occupancy will be significantly smaller than the other dopants and still it has enough magnetic moment to destroy the time-reversal symmetry. In $Bi_2Se_3$, the maximum valency of Bi is (+3), so if, we dope Ti at Bi site, it is expected to have about +3 valency; homovalent substitution. Hence, it is a good material to show weak magnetism and its role in the topological behavior. $Bi_2Se_3$ forms in rhombohedral crystal structure with space group $D_{3d}^5$ ($R\bar{3}m$). Reported lattice-parameters are $a = b = 4.143$ Å, $c = 28.636$ Å, $\alpha = \beta = 90^0$, $\gamma = 120^0$. $Bi_2Se_3$ crystal structure is shown in Fig.1. Two in-equivalent Se atoms are indexed with Se1 and Se2. The whole crystal is composed of Bi and Se layer in a repetitive manner. Our main aim was to prepare good quality single crystals of $Bi_{2-x}Ti_xSe_3$ for small values of $x$ and characterize the samples using various characterization techniques. Small Ti substitutions will allow to dope magnetic impurities without significant change in crystal structure.

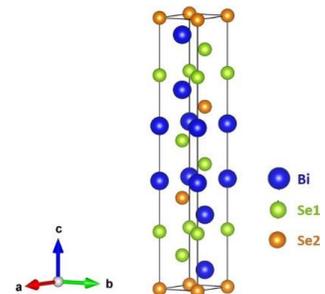

**FIGURE 1.** Unit cell of $Bi_2Se_3$

## EXPERIMENT

We have prepared the single crystal of $Bi_2Se_3$ and its Ti-doped compounds using the Bridgman technique. $Bi_2Se_3$ melts congruently, this type of material can be easily formed by the Bridgman method. To get a good quality single crystal, we cooled down the sample from 705 °C down to room temperature very carefully. In the furnace, we set the temperature profile shown in Fig.2. The melting point of Bi and Se are 271 °C and 221 °C, respectively. Melting point of Ti is quite high 1668 °C but it is soluble in molten Se. To get a homogeneous mixture of (Bi, Se) and Ti, we kept this solution at 900 °C temperature for 1 day in the furnace. Then the furnace was cooled at a rate of 100 °C/hr. to 725 °C. Below 705 °C, solidification starts, so we cool down at a smaller rate i.e. (1 °C/hr) up to temperature 675 °C, followed by 60 °C/hr, down to room temperature. We have prepared $Bi_2Se_3$, $Ti_{0.01}Bi_{1.99}Se_3$, $Ti_{0.03}Bi_{1.97}Se_3$ and $Ti_{0.10}Bi_{1.90}Se_3$. Photoemission measurements were carried out in a spectrometer equipped with Phoibos 150 electron detector from Specs GmbH, Germany and monochromatic Al K$\alpha$ x-ray source. The samples were cleaved in situ before doing the measurements. We confirmed the purity of the surface before measurements by monitoring O 1$s$ and C 1$s$ signals in the survey scans.

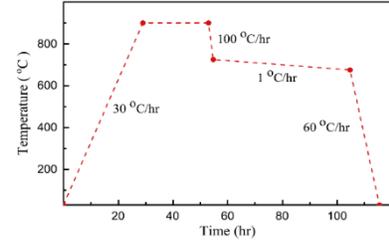

**FIGURE 2.** Time-temperature profile of crystal growth of $Bi_2Se_3$

## RESULTS AND DISCUSSIONS

We have done the energy dispersive analysis of $x$-ray on several pieces of a particular sample with an electron beam of kinetic energy 20 keV. The mean values of each sample are tabulated below in Table I. We found that there is significant Se deficiency in all samples but a significant amount of Ti is present in every sample and it substituted Bi site.

**TABLE 1.** EDX analysis Data, all results are in atomic percentage

| Sample | Ti | Bi | Se |
|---|---|---|---|
| $Bi_2Se_3$ | 0.00 | 43.37 | 56.63 |
| $Ti_{0.01}Bi_{1.99}Se_3$ | 0.18 | 42.65 | 57.17 |
| $Ti_{0.03}Bi_{1.97}Se_3$ | 0.61 | 40.37 | 59.02 |
| $Ti_{0.10}Bi_{1.90}Se_3$ | 2.14 | 39.89 | 57.97 |

X-ray diffraction was done using monochromatic Cu $K\alpha$ radiation of wavelength 1.5418 Å. To find the lattice parameters, we have used the FullProf software package. All lattice parameters are tabulated in Table. II, which clearly shows that the $c$ axis gets contracted due to Ti doping as its atomic radius is small compared to the atomic radius of Bi.

**TABLE 2.** Lattice parameters

| Sample | a (Å) | b (Å) | c (Å) | $\alpha$ (°) | $\beta$ (°) | $\gamma$ (°) |
|---|---|---|---|---|---|---|
| $Bi_2Se_3$ | 4.140 | 4.140 | 28.649 | 90 | 90 | 120 |
| $Ti_{0.03}Bi_{1.97}Se_3$, | 4.141 | 4.141 | 28.642 | 90 | 90 | 120 |
| $Ti_{0.10}Bi_{1.90}Se_3$ | 4.141 | 4.141 | 28.630 | 90 | 90 | 120 |

In Fig.3, we show the powder $x$-ray diffraction patterns of the three samples. Miller indices of the planes corresponding to each peak are written in the figure.

Comparing the diffraction patterns, we can see that the position of each peak of $Bi_2Se_3$ and $Ti_{0.03}Bi_{1.97}Se_3$ are almost same but there is a shift to a higher value for $Ti_{0.10}Bi_{1.90}Se_3$. As mentioned before, Ti is smaller in size compared to Bi leading to a decrease in inter-planar distance with doping of Ti at Bi site. Hence for a particular value of $n$ and $\lambda$, $\theta$ will increase (Bragg's law $2d\sin\theta = n\lambda$). So, the peak position will shift to higher theta value as can be seen in the data for $Ti_{0.10}Bi_{1.90}Se_3$. For $Ti_{0.03}Bi_{1.97}Se_3$, the amount of Ti is too small to make a visible shift. These results establish the good quality of the sample.

We have carried out x-ray photoemission spectroscopy measurements on a freshly cleaved surface of $Bi_2Se_3$ and $Ti_{0.10}Bi_{1.90}Se_3$ to investigate the composition and the change in the electronic structure of $Bi_2Se_3$ with Ti doping via analysis of the core level spectra. We used Al $K\alpha$ (photon energy $h\nu$ =1486.6 eV) as x-ray source and the experiment was done at 88 K for $Bi_2Se_3$ and room temperature for Ti-doped $Bi_2Se_3$ in an ultra-high vacuum condition (base pressure $10^{-11}$ Torr). It is to note here that although Ti-doped sample could not be measured at low temperature, temperature does not affect the electronic structure apart from slight broadening at higher temperatures due to phonon excitations. The second concern is the energy resolution – Ti-doped sample is measured with higher pass energy due trouble with the x-ray gun and hence the energy resolution for the Ti-doped case is close to 0.6 eV while it is 0.35 eV in the case of $Bi_2Se_3$. We have taken the data for normal emission of photoelectrons. The survey scans of the freshly cleaved samples do not show signature of any impurity and a sufficient amount of Ti 2p signal is present in the spectra. These observations ensures the purity of the sample prepared.

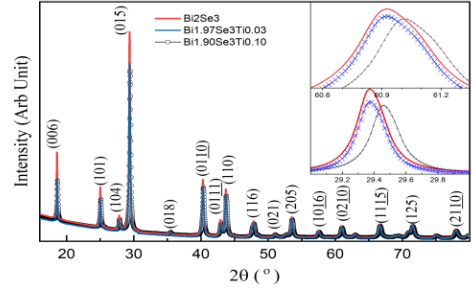

**FIGURE 3.** Powder x-ray diffraction pattern of $Bi_2Se_3$ and Ti doped $Bi_2Se_3$

Bi 4d core level spectrum of $Bi_2Se_3$ is shown Fig.4 (a). Spin-orbit split Bi $4d_{5/2}$ and Bi $4d_{3/2}$ are located at 441.1 eV and 464.9 eV binding energies exhibiting a spin-orbit splitting of 23.8 eV. Features around 460.5 eV and 482.7 eV are plasmon loss features, which appear due to collective excitations of charge-density oscillation[1]. Along with the plasmon loss feature, we observe two relatively less intense features around binding energies 448.3 eV and 472.3 eV; the energy difference between their peak positions is almost equal to the spin-orbit splitting of Bi 4d features (23.8 eV). These features may be attributed to the satellite signals due to final state effect. It is to note here that the formation of satellite depends on how the electrons are correlated. Since, Bi 6p electrons are known to be less influenced by electron correlation, such features are not expected, in general.

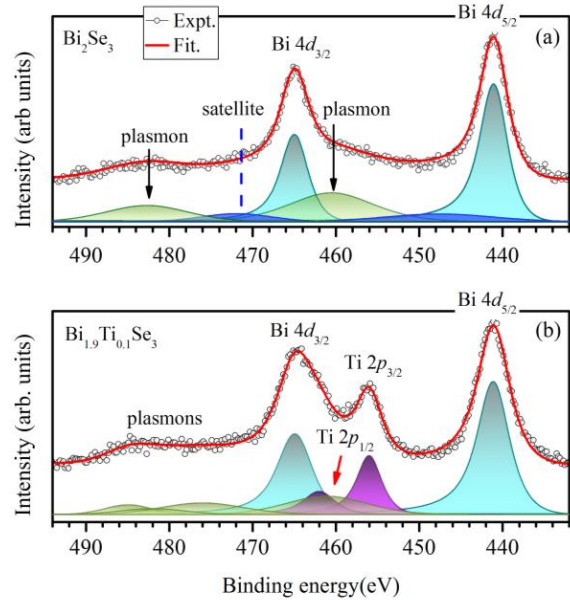

In Fig. 4(b), we show the photoemission spectra of $Ti_{0.10}Bi_{1.90}Se_3$. In this case, the energy resolution broadening is larger than the $Bi_2Se_3$ case for the reason discussed above. Distinct signature of Ti 2p signal is observed at 458 eV along with the Bi 4d peaks. Photoemission signal corresponding to Ti $2p_{1/2}$ is hidden with Bi $4d_{3/2}$ peak; the spin orbit splitting of Ti 2p is estimated to be close to 6 eV [4]. The higher energy region can be simulated with 4 plasmon peaks with an energy separation of 20 eV from the main peak; the additional plasmon peaks are associated with two Ti 2p peaks. Due to presence of many distinct broad features, it was difficult to simulate the satellite features due to final state effect with certainty and hence not shown here.

**FIGURE 4.** Bi 4d photoemission spectra of (a) $Bi_2Se_3$ and (b) Ti doped $Bi_2Se_3$. Distinct signature of Ti 2p signals are also seen in the figure.

In Fig. 5, we show the Bi 4f spectra collected from both $Bi_2Se_3$ and $Bi_{1.9}Ti_{0.1}Se_3$. The data exhibit spin-orbit split Bi $4f_{7/2}$ and Bi $4f_{5/2}$ peaks at 158 eV and 163.3 eV binding energies, respectively; spin-orbit splitting is about 5.3 eV. Around 160 eV and 165.8 eV, two peaks of very low intensity are seen due to Se 3p excitations [1]. The spectrum collected from Ti doped compound is shown in Fig. 5 (b). Due to large resolution broadening, the sharpness of the features seen in Fig. 5(a) is not present in Fig. 5(b). However, distinct signal of Bi 4$f$ spin-orbit split features are observed in the spectra. The large width smeared out the signature of Se 3$p$ signals too. It is to be noted here that the additional broadening observed due to instrumental resolution appear to be smaller in the case of Bi 4$d$ than Bi 4$f$ – this is primarily due to the energy scale effect. Intrinsic effect such as Ti substitution induced disorder, if there is any, is not visible in the presented data. Further study with better energy resolution is necessary.

## CONCLUSIONS

In summary, we could successfully prepare a very high quality Bi$_2$Se$_3$ single crystal and its Ti-doped compositions using Bridgman technique. All XRD and XPS data of Bi$_2$Se$_3$ show a nice agreement with previously reported results with better quality of the experimental spectra. X-ray diffraction pattern of the doped compositions exhibit almost identical patterns with small shift in the diffraction peak positions. This indicates that Ti actually goes to the intended substitution sited.

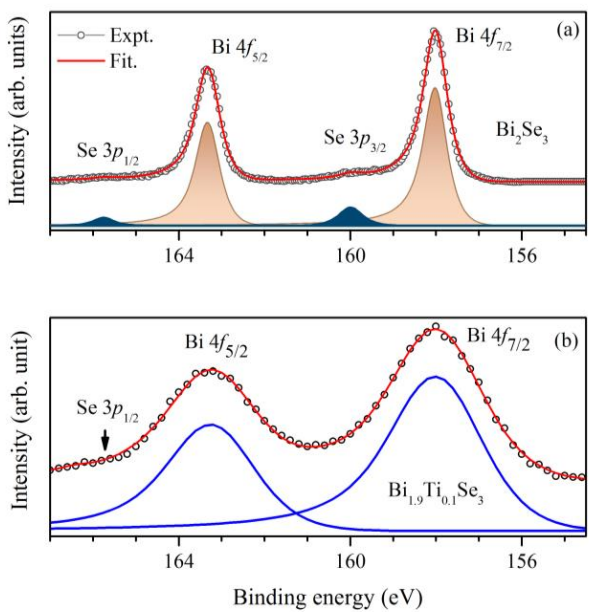

**FIGURE 5.** Bi 4$f$ photoemission spectra of (a) Bi2Se3 and (b) Bi$_{1.9}$Ti$_{0.1}$Se$_3$.

The analysis of the crystal structure exhibit a small decrease in lattice parameters with the increase in Ti concentrations due to lattice contraction arising from smaller ionic radius of Ti relative to the radius of Bi. The photoemission spectra exhibit distinct and profound signal of Ti 2$p$; the spectral broadening observed here primarily due to resolution broadening. In addition, the core level spectra exhibit signature of satellites due to plasmon excitations and photoemission final state effects even in the parent compound Bi$_2$Se$_3$. Since, the Coulomb repulsion strength among Ti 3$d$ electrons is strong, substitution of Ti at Bi site is expected to enhance the correlation effect. Such study requires photoemission measurements with hign energy resolution.

## ACKNOWLEDGMENTS

SD thanks Dr. Rajib Mondal for his help during the sample preparation. KM acknowledges financial assistance from the Department of Science and Technology, government of India under J. C. Bose Fellowship program and the Department of Atomic Energy under the DAE-SRC-OI Award program.